\documentstyle{cupconf}

\ifoldfss
\else
  \ifnfssone
    \newmathalphabet{\mathit}
      \addtoversion{normal}{\mathit}{cmr}{m}{it}
      \addtoversion{bold}{\mathit}{cmr}{bx}{it}
    \newmathalphabet{\mathcal}
      \addtoversion{normal}{\mathcal}{cmsy}{m}{n}
    \else
    \ifnfsstwo
    \fi
  \fi
\fi

\def\'#1{\ifx#1i{\accent"13\i}\else{\accent"13#1}\fi}
\def\BP{Ballesteros-Paredes}
\def\gamef{\gamma_{\rm e}}
\def\gtsima{$\; \buildrel > \over \sim \;$}    
\def\ltsima{$\; \buildrel < \over \sim \;$}    
\def\M{M_{\rm rms}}
\def\Peq{P_{\rm eq}}
\def\rhot{\rho_{\rm t}}
\def\simgt{\lower.5ex\hbox{\gtsima}}           
\def\simlt{\lower.5ex\hbox{\ltsima}}           

\def\VS{V\'azquez-Semadeni}

\def\hexnumber#1{\ifcase#1 0\or1\or2\or3\or4\or5\or6\or7\or8\or9\or
 A\or B\or C\or D\or E\or F\fi }

\makeatletter
\ifx\CUP@mtlplain@loaded\undefined
\else
\fi
\makeatother
 \makeatletter
 \ifx\CUP@mtlplain@loaded\undefined
   \font\tenbmi=cmmib10 at 10pt
   \font\sevenbmi=cmmib10 at 7pt
   \font\fivebmi=cmmib10 at 5pt

   \newfam\bmifam
   \textfont\bmifam=\tenbmi
   \scriptfont\bmifam=\sevenbmi
   \scriptscriptfont\bmifam=\fivebmi
   
 \fi
 \makeatother
\ifnfsstwo

\fi
\ifnfssone

\fi
\ifoldfss

\fi

\mathchardef\varLambda="0103

\makeatletter
\ifx\CUP@mtlplain@loaded\undefined
\else
\fi
\makeatother
\makeatletter
\ifx\CUP@mtlplain@loaded\undefined
  \font\tenbms=cmbsy10
  \font\sevenbms=cmbsy10 at 7pt
  \font\fivebms=cmbsy10 at 5pt
  \newfam\bmsfam
  \textfont\bmsfam=\tenbms
  \scriptfont\bmsfam=\sevenbms
  \scriptscriptfont\bmsfam=\fivebms

  \edef\bsy@{\hexnumber\bmsfam}
  \mathchardef\bnabla="0\bsy@72
\fi
\makeatother
\title[Turbulence as Organizing Agent in ISM]{Turbulence as an
Organizing Agent in the ISM}

\author[\VS\ \& Passot]%
{E\ls N\ls R\ls I\ls Q\ls U\ls E\ns  V\ls A\ls Z\ls Q\ls U\ls \ls E
Z\ls -\ls S\ls E\ls M\ls A\ls D\ls E\ls N\ls I$^1$\\ 
\and\ns T\ls H\ls I\ls E\ls R\ls R\ls Y\ns  P\ls A\ls S\ls S\ls O\ls T$^2$}

\affiliation{$^1$Instituto de Astronom\'ia, UNAM, Apdo.\ Postal
70-264, M\'exico, D.\ F.\ 04510, MEXICO\\[\affilskip]
$^2$Observatoire de la C\^ote d'Azur, B.P.\ 4229, 06304, Nice
  Cedex 4, FRANCE}

\begin{document}
\ifnfssone
\else
  \ifnfsstwo
  \else
    \ifoldfss
      \let\mathcal\cal
      \let\mathrm\rm
      \let\mathsf\sf
    \fi
  \fi
\fi

\maketitle

\begin{abstract}
We discuss HD and MHD compressible turbulence as a cloud-forming and
cloud-structuring mechanism in the ISM. Results from a numerical model
of the turbulent ISM at large scales suggest that the phase-like
appearance of the medium, the typical values of the densities
and magnetic field strengths in the intercloud medium, as
well as the velocity dispersion-size scaling relation in clouds may be
understood as consequences of the interstellar turbulence. However,
the density-size relation appears to only hold for the densest clouds,
suggesting that low-column density clouds, which are hardest to
observe, are turbulent transients. We then explore some properties of
highly compressible polytropic turbulence, in one and several
dimensions, applicable to molecular cloud scales. At low values of the
polytropic index $\gamma$, turbulence may induce the gravitational
collapse of otherwise linearly stable clouds, except if they are
magnetically subcritical. The nature of the density fluctuations in
the high Mach-number limit depends on $\gamma$, and in no case
resembles that resulting from Burgers turbulence. In the isothermal
($\gamma=1$) case, the dispersion of $\ln (\rho)$ scales like the
turbulent Mach number. The latter case is singular with a
lognormal density pdf, while power-law tails develop at high
(resp. low) densities for $\gamma <1$ (resp. $\gamma >1$).
\end{abstract}

\firstsection 
\section{Introduction}

One of the main features of turbulence is its multi-scale nature
(e.g., \cite{scalo87}; \cite{lesieur90}). In particular, in the
interstellar medium (ISM),
relevant scale sizes span nearly 5 orders of magnitude, from the size of
the
largest complexes or ``superclouds'' ($\sim 1$ kpc) to that of dense
cores in molecular clouds (a few $\times 0.01$ pc), with densities
respectively ranging from $\sim 0.1$ cm$^{-3}$ to $ \simgt 10^6$
cm$^{-3}$. Moreover, in the diffuse gas itself, even smaller scales,
down to sizes several $\times 10^2$ km are active (see the chapters by
Spangler and Cordes), although at small densities. Therefore,
in a unified turbulent picture of the ISM,
it is natural to expect that turbulence can intervene in
the process of cloud formation (\cite{hunter79}; \cite{hunt_fleck82};
\cite{elm93}; \VS, Passot \& Pouquet 1995, 1996) through modes larger
than the clouds 
themselves, as well as in providing cloud support and determining the
cloud properties, through modes smaller than the clouds
(\cite{chandra51}; \cite{bonazzola87}; \cite{Leorat90};
\cite{VS_gaz95}). Moreover, 
another essential feature of turbulence is that all these scales
interact nonlinearly, so that coupling is expected to exist between
the large-scale cloud-forming modes and the small-scale cloud
properties.

In this chapter we adopt the above viewpoint as a framework for
presenting some of the most relevant results we have learned from
two-dimensional (2D) numerical simulations of the turbulent ISM in a
unified and coherent 
fashion, as it relates to the problems of cloud formation, the
phase-like structure of the ISM and the
topology of the magnetic and density fields, as well as internal cloud
properties, such as their virialization and scaling relations (\S\
\ref{2D}). Next
we discuss recent results from multi-dimensional simulations and a
simple heuristic 
model of one-dimensional polytropic turbulence, as a first attempt to
gain more physical insight into the mechanisms responsible for the
generation of the statistics of the density fluctuations in
compressible turbulence (\S\ \ref{polytropes}). Finally, we present a
summary and conclusions in \S\ \ref{conclusions}. 

\section{Cloud Formation and Properties in the Turbulent ISM} \label{2D}

In a series of recent papers (\VS\ et al.\ 1995 (Paper I), 1996 (Paper
III); Passot, \VS\ \& Pouquet 1995 (Paper II)), we have
presented two-dimensional (2D) numerical simulations of
turbulence in the ISM on the Galactic plane, including self-gravity,
magnetic fields, simple parametrizations 
of standard cooling functions (\cite{dalg_McCray72};
\cite{raymond_etal76}) as given by \cite{chiang_breg88}, diffuse
heating mimicking that of background UV radiation and cosmic rays,
rotation, and a simple prescription for star formation (SF)
which represents massive-star ionization heating by turning on a local
source of heat wherever the density exceeds a threshold
$\rhot$. Supernovae are now being included (Ga\-zol-Pa\-ti\-\~no \&
Passot 1998; see also Korpi, this 
Conference, for analogous simulations in 3D). The
simulations follow the evolution of a 1 kpc$^2$ region of the ISM at
the solar Galactocentric distance over $\sim 10^8$ yr and are started
with Gaussian fluctuations with random 
phases in all variables. The initial fluctuations in the velocity
field produce shocks which trigger star formation which, in turn, feeds
back on the turbulence, and a self-sustaining cycle is
maintained. These simulations have been able to reproduce a number of
important properties of the ISM, suggesting that the processes
included are indeed relevant in the actual ISM. Some interesting
predictions have also resulted.

\subsection{Effective Polytropic Behavior and Phase-Like Structure}
\label{phases}

\begin{figure}   
\vspace{15pc}
\begin{minipage}{15pc}
\includegraphics{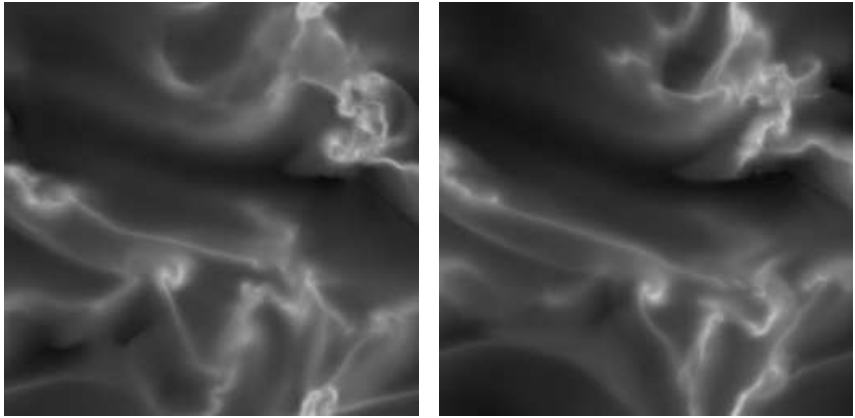}
\end{minipage}
  \caption{Two views of the density field in a simulation of 1 kpc$^2$
of the ISM on the Galactic plane, one at a) $t=6.37\times10^7$ yr
(left), the other at b) $t=7.80\times 10^7$ yr (right). The large-scale
structures are contracting gravitationally, while the smaller structures
within them change significantly due to the turbulent motions.}
\label{collapse}
\end{figure}

One of the earliest results of the simulations is a consequence of the
rapid thermal rates (\cite{spit_sav50}), faster than the dynamical
timescales by factors of 10--$10^4$ in the simulations (Paper
I). Thus, the gas is
essentially always in thermal equilibrium, except in star-forming
regions, and an effective
polytropic exponent $\gamef$ (\cite{elm91}) can be calculated,
which results from 
the condition of equilibrium between cooling and diffuse heating,
giving an effectively polytropic behavior $\Peq \propto
\rho^{\gamef}$, where $\rho$ is the gas density (see 
Papers II and III for details). Even though the heating and cooling
functions used do not give a thermally unstable (e.g.,
\cite{field_etal69}; \cite{balbus95})
regime at the temperatures reached by the simulations, they manage to
produce values of $\gamef$ smaller than unity for temperatures in the
range 100 K $< T < 10^5$ K, implying that {\it denser regions are
cooler}. Upon the production of turbulent density
fluctuations, the flow reaches a temperature distribution similar to
that resulting from isobaric thermal instabilities (\cite{field_etal69}),
but without the need for them. Note, however,
that in this case there are no sharp phase transitions.

\subsection{Cloud Formation}\label{formation}

In the simulations, the largest cloud complexes (several hundred pc)
form simply by gravitational instability. Although in Paper I it was
reported that no gravitationally bound structures were formed, this
conclusion did not take into account the effective reduction of the
Jeans length due to the small $\gamef$ of the fluid. Once this effect
is considered, it is found that the largest scales in the simulations
are unstable. This process is illustrated in fig.\ \ref{collapse},
which shows two snapshots of the logarithm of the density in a
simulation at a resolution of 512 grid 
points per dimension (run 28 from Paper II), one at $t=6.37\times
10^7$ yr (a), with minimum and maximum densities of 0.04 and 36.1
cm$^{-3}$, and the other at $t=7.80\times 10^7$ yr, with extrema of
0.046 and 58 cm$^{-3}$. The two very large scale structures in
the upper and lower halves of the integration box, are seen to have
contracted at the later time, and the voids have
expanded. Nevertheless, inside such large-scale clouds, an extrememly
complicated morphology is seen in the higher-density material, as a
consequence of the turbulence generated by the star
formation activity. The medium- and small-scale clouds are
thus turbulent density fluctuations.

\subsection{Cloud and magnetic field topology}\label{topology}

\begin{figure} 
  \vspace{15pc}
\begin{minipage}{15pc}
\includegraphics{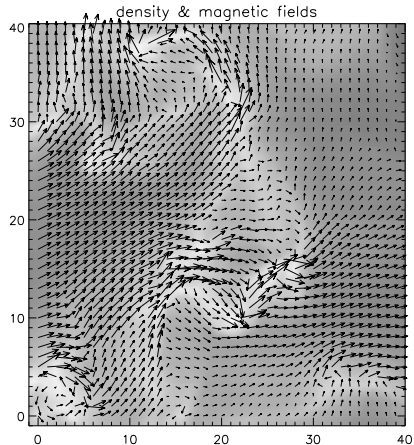}
\end{minipage}
  \caption{Gray-scale image of the logarithm of the density field,
with superimposed magnetic field vectors. Shown is a subfield of $200
\times 200$ pc ($160 \times 160$ pixels), from a simulation at
resolution of 800 grid points per dimension ({VBR97}). The minimum and
maximum magnetic field intensities are 0.13 and 26.6 $\mu$G,
respectively. The axes show arbitrary units. See text for feature
description.}
\label{dens_mag}
\end{figure}

The topology of the clouds formed as turbulent fluctuations in the
simulations is extrememly filamentary. This property apparently
persists in 3D simulations (see chapters by Ostriker, Stone, Mac Low and
Padoan). Interestingly, the magnetic field also
exhibits a morphology 
indicative of significant distortion by the turbulent motions (Paper
II; \cite{BPVS98}). The field has a tendency to be aligned with density
features, as shown in fig.\ \ref{dens_mag}. Even in the presence of 
a uniform mean field, motions along the latter amplify the perpendicular
fluctuations due to flux freezing, while at the same time they produce
density fluctuations elongated perpendicular to the direction of
compression. This mechanism also causes many of the density
features to contain 
magnetic field reversals (e.g., the feature near the lower left corner)
and bendings (e.g., the feature down and to the right off the
center, with coordinates $x \sim 25, y\sim 12$). It happens also that
magnetic fields can traverse the clouds 
without much perturbation, as seen for example in 
the feature at $x=21, y=30$. These results are consistent with the
observational result that the magnetic field does not seem to vary
much along clouds (\cite{goodman_etal90}), and in general does not
present a unique kind of alignment with the density features. On the
other hand, recent observations have found field bendings similar to
those described here (Crutcher, this volume). Also, field
reversals in clouds have been recently observed (\cite{heiles97}).

It is important to note that the ``pushing'' of the turbulence
on the magnetic field occurs for realistic
values of the energy injection from stars 
and of the magnetic field strength, which
ranges from $\sim 5 \times 10^{-3} \mu$G (occurring at the low density
intercloud 
medium) to a maximum of $\sim 25 \mu$G, which occurs in one of the
high density peaks, although with no unique $\rho$-$B$ correlation
(\cite{paperII}). Observationally, larger values of the field occur
only on much smaller scales than those resolved by our simulations
(1.25 pc in at resolution of $800^2$ grid points)
(\cite{heiles_etal93}). Thus, the 
simulations suggest that the effect of the magnetic field is not as
strongly dominating as often assumed in the literature.
This is also in agreement with
the fact that the magnetic and kinetic energies in the simulations are
in near global equipartition at all scales, as shown by their energy
spectra (fig.\ 5 in \cite{paperII}).

Finally, note that the fact that the magnetic spectrum
exhibits a clear self-similar (power-law) range, together with the
fact that the fluctuating component of the field is in general
comparable or larger than the uniform field, suggests strongly that
the medium is in a state of fully developed MHD turbulence, rather than
being a superposition of weakly nonlinear MHD waves.

\subsection{Cloud scaling properties}\label{scaling}

An important question concerning the clouds formed in the simulations
is whether they reproduce some well-known observational scaling and
statistical properties of interstellar clouds, most notably the
so-called Larson's relations between velocity dispersion $\Delta v$,
mean density $\rho$ and size $R$ (\cite{larson81}), and the cloud mass
spectra (e.g., \cite{blitz91}). \VS, \BP\ \& Rodr\'iguez (1997,
hereafter \cite{VBR97}) have
studied the scaling properties of the clouds in 
the simulations, finding that the cloud ensemble exhibits a relation
$\Delta v \propto R^{0.4 \pm 0.08}$ and a cloud mass spectrum $dN(M)/dM
\propto M^{-1.44 \pm 0.1}$, both being consistent with observational
surveys, especially those specifically including gravitationally
unbound objects (e.g., \cite{falg_etal92}). However, it was found that
no density-size relation like that of Larson ($\rho \propto R^{-1}$)
is satisfied by the clouds in the simulations. Instead, the clouds
occupy a triangular region in a $\log \rho$--$\log R$ diagram, as
shown in fig.\ 8 of \cite{VBR97}, with only its upper envelope being close
to
Larson's relation. This implies the
existence of clouds of very low column density, which are presumably
turbulent transients, and can be easily missed by observational surveys
if they do not integrate for long enough times. A few observational
works, however, point towards the existence of transients
(\cite{loren89}; \cite{magnani_etal93}) and low-column density
clouds, with masses much smaller than those estimated from virial
equilibrium (\cite{falg_etal92}).

An implication of Larson's relations is the so-called logatropic
``equation of state'' (\cite{liz_shu89}). \cite{VS_etal98}
have investigated wheth\-er this behavior is verified in numerical
simulations of gravitational collapse with initially turbulent
conditions. A logatropic behavior would imply a scaling $\Delta v
\propto \rho^{-1/2}$. However, a scaling $\Delta v \propto
\rho^\alpha$, with $1/4 <\alpha < 1/2$ was observed, suggesting a
polytropic
behavior instead. This was interpreted as meaning that the logatropic
equation of state was obtained by an invalid assumption, namely that
Larson's relations 
are applicable to a thermodynamic process on a cloud of fixed
mass. Instead, they seem to represent only the conditions of a
(possibly relaxed)  ensemble of
clouds of different masses, and so are inapplicable to the former
case.

\section{Results on polytropic compressible turbulence} \label{polytropes}

\subsection{Production and stability of turbulent density
fluctuations} \label{poly_general}

In view of the effective polytropic behavior exhibited by the
simulations (\S\ \ref{phases}), a natural abstraction is
to consider the behavior of purely polytropic fluids, whose equation
of state is $P=\rho^{\gamef}/\gamef$. For such a fluid, it has been
shown in (\cite{paperIII}) that the density jump $X\equiv \rho_2/\rho_1$
in a shock in a polytropic gas satisfies
\begin{equation}
X^{1+\gamef} - (1+ \gamef M^2)X + \gamef M^2=0,
\end{equation}
where $M$ is the Mach number upstream of the shock. From this
equation, we recover the fact that the compression ratio for an
isothermal shock ($\gamef =1$) is $M^2$, but we also see that $X \rightarrow 
e^{M^2}$ as $\gamef \rightarrow 0$, a density jump which can be much
larger than the isothermal one.

Turbulence-induced fluctuations can either collapse or rebound,
depending on their cooling and dissipating abilities
(e.g., \cite{hunt_fleck82}; \cite{hunt_etal86}; Elmegreen \&\
Elmegreen 1978;
Vishniac 1983, 1994; \cite{elm93}). The critical value of $\gamef$ for
which a turbulent density fluctuation formed by an 
$n$-dimensional compression can collapse was given in \cite{paperIII} as
$\gamma_{\rm cr} \equiv 2(1-1/n)$. The same criterion
was given by \cite{mckee_etal93} for fixed $\gamef \sim 1$ as an indication
that 1-dimensional shock compressions cannot cause collapse. Instead,
we can argue that the combination of small enough $\gamef$ and
compressions in more than one dimension (shock collisions) can trigger
collapse. \cite{scalo_etal98} have recently discussed the possible
values of $\gamef$ in the cold ISM, finding that, although with large
uncertainty, $\gamef \sim 1/3$ is possible at densities $\simgt 5
\times 10^4$ cm$^{-3}$, thus making the collapse of shock-compressed
cores feasible.

\subsection{Statistics of density fluctuations in polytropic
turbulence} \label{pdfs}

The turbulent formation of clouds in the ISM must ultimately be
described by the statistics of density fluctuation
production in compressible turbulence. Therefore, it is of interest to
investigate the probability density function (pdf) of the density
fluctuations that develops in numerical simulations. Interestingly,
the pdfs reported for a variety of flows show important qualitative
differences. \cite{porter_etal91} reported an exponential pdf for 3D,
weakly compressible thermodynamic turbulence. Power-law pdfs have been
reported for low-Reynolds number, one-dimensional Burgers flows
(\cite{gotoh_kraich93}) and for the simulations described in \S\
\ref{2D}, as well as for two-dimensional Burgers flows
(Scalo et al.\ 1998), while lognormal pdfs have been reported for
isothermal 2D (\cite{VS94}) and 3D (\cite{padoan_etal97}) simulations.

Passot \& V\'azquez-Semadeni (1998, hereafter \cite{PVS98}) have
investigated the simplest case of a 
one-dimensional purely hydrodynamic flow by means of a heuristic model
and very high resolution (up to 6144 grid points) numerical
simulations. Here, we summarize these results 
briefly. In order to study the production of the local density
fluctuations, it is convenient to describe them
as a sequence of isolated, discrete jumps
(\cite{VS94}). Consider first the isothermal ($\gamef=1$) case,
whose governing equations read
\begin{eqnarray}
\frac{Du}{Dt} =-\frac{1}{M^2}\frac{\partial}{\partial
  x}s \label{eq:Su} \\
\frac{D s}{Dt} = - \frac{\partial}{\partial  x} u,  \label{eq:Sr}
\end{eqnarray}
where $\rho$ is the fluid density, $u$ is the velocity, $s=\ln \rho$
and $M$ is the Mach number of the velocity unit.
Note that these equations are invariant upon the change
$s\rightarrow s+b$, where $b$ is an arbitrary constant, reflecting the
fact that the sound speed does not depend on the density in this
case. Consider now the sequence of density jumps $\rho_2/\rho_1$. This
is a sequence of multiplicative steps, and, consequently, additive
steps in $s=\ln \rho$. But, because of the translation invariance
mentioned above, a jump of a given magnitude must have the same
probability of occurrence, independently of the initial density. Thus,
the sequence involves events with the same probability distribution,
and by the Central Limit Theorem it must converge to a Gaussian
distribution in $s$, or, equivalently, a lognormal distribution in
$\rho$ (see Nordlund, this volume, for an alternative derivation),
explaining the reported pdfs in the isothermal case.

\begin{figure} 
  \vspace{15pc}
\begin{minipage}{15pc}
\includegraphics{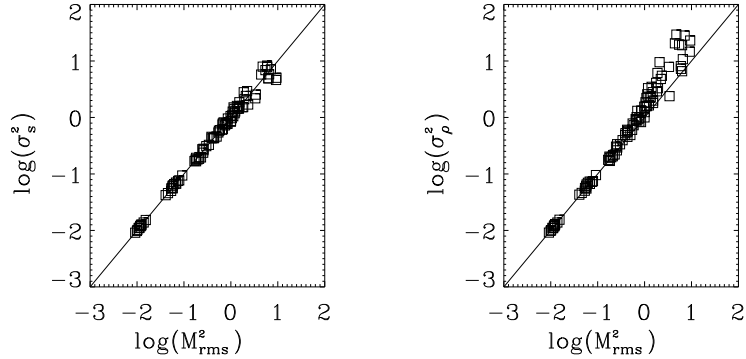}
\end{minipage}
  \caption{Logarithm of the variance of a) $s \equiv \ln \rho$ (left)
and b) the density $\rho$ (right) vs.\ the logarithm of the rms Mach
number in one-dimensional simulations of
polytropic turbulence. Note that $\sigma_s^2$ scales as $M_{\rm rms}^2$,
while $\sigma_\rho^2$ increases faster than $\M^2$.}
\label{sigma_g1}
\end{figure}

In order to fully characterize the pdf, it is necessary to determine
its mean and variance. Concerning the latter, \cite{PVS98} have suggested,
through an anlysis of the shock and expansion waves in the system, that
for a large range
of Mach numbers the typical size of the logarithmic jump is expected to
be $\sigma_s \sim \M$, where $\M$ is the
rms Mach number. This result is
verified numerically, as shown in fig.\ \ref{sigma_g1}a. Note that
fig.\ \ref{sigma_g1}b shows the scaling of the linear density
variance $\sigma_{\rho}^2$ vs.\ $\M^2$. An exponential
behavior is observed, most noticeable at large Mach numbers, in
agreement with the 
relation $\sigma_s^2 = \ln (1 + \sigma_{\rho}^2)$ which holds for a
lognormal distribution. This contrasts with recent claims that it is
$\sigma_\rho$ which scales as $\M$ (\cite{padoan_etal97}; Nordlund, this
volume). This discrepancy can be understood as a consequence of the
similarity between the two variances at small Mach 
numbers, together with the fact that the simulations from which those
authors
reached their conclusion were three-dimensional, implying that a
significant fraction of the kinetic energy was in rotational modes
(reportedly $\sim 80$ \%), and thus not available for producing
density fluctuations. Instead, in the 1D simulations of \cite{PVS98},
all of the kinetic energy is compressible.

Concerning the mean $s_0$ of the distribution, it can be directly
evaluated from the mass conservation condition $\langle \rho \rangle
=\int_{-\infty}^{+\infty} e^s P(s) ds = 1$, where $P(s)$ is the
pdf of $s$, yielding $s_0=-\sigma_s^2/2$. With all this in mind, we
can then write the model pdf for $s$ as
\begin{equation}
P(s)ds=\frac{1}{\sqrt{2\pi \sigma_s^2}}\exp\bigl[-\frac{(s-s_o)^2}
{2\sigma_s^2}\bigr] ds,
\end{equation}
with $\sigma_s^2=\beta \M^2$, and $\beta$ a proportionality
constant.  The numerical simulations confirm the dependence of the
width and mean of the distribution with $\M$ (\cite{PVS98}).

We next consider the case $\gamef \ne 1$. The governing equations
can now be written as
\begin{eqnarray}
\frac{Du}{Dt} =\frac{1}{(1-\gamef)M^2}\frac{\partial}{\partial
  x}e^{-v}\label {eq:Sigu} \\
\frac{Dv}{Dt} = -(1-\gamef) \frac{\partial}{\partial  x} u.
\label{eq:Sigr}
\end{eqnarray}
Interestingly, these equations return to the form of eqs.\ (\ref{eq:Su})
and (\ref{eq:Sr}) upon the density-dependent rescaling $M\rightarrow
M(s;\gamef)= Me^{(1-\gamef)s/2}$. Thus, we formulate the ansatz that
the form of the pdf also remains the same, provided the above
replacement is made. After relocating the  term in $s_0$ from inside
the exponential function to the normalization constant, we can write
the model pdf as 
\begin{equation}
P(s;\gamef)ds=C(\gamef) \exp \Bigl[\frac{-s^2
e^{(\gamef-1)s}}{2M^2} - \alpha(\gamef)s\Bigr] ds. \label{eq:PDFgne1}
\end{equation}
\begin{figure} 
  \vspace{15pc}
\begin{minipage}{15pc}
\includegraphics{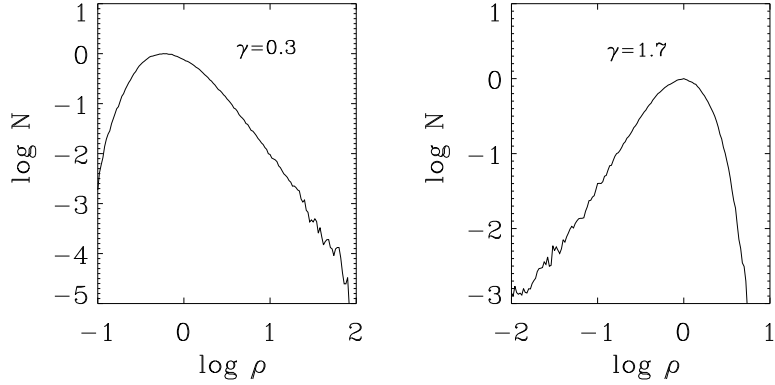}
\end{minipage}
  \caption{Density pdfs for two 1D simulations of polytropic
turbulence, one with $\gamef=0.3$ (left), and the other with
$\gamef=1.7$ (right), both with $M=3$.}
\label{pdfs_gne1}
\end{figure}
Note that this is a particular form of the pdf valid only in a range
of $s$-values (\cite{PVS98}), but for illustrative purposes it
suffices here. This equation shows that when $(\gamef-1)s<0$, the pdf
asymptotically approaches a power law, while in the opposite case it
decays faster than a lognormal. Thus, for $0 <\gamef <1$, the pdf
approaches a power law at large densities ($s>0$), and at low
densities for $\gamef >1$. The former case is in agreement with the
pdfs reported 
for our 2D simulations, which in general have $\gamef<1$. The 1D
simulations also verify this result for $\gamef>1$ (fig.\
\ref{pdfs_gne1}). See Nordlund (this volume) for a parallel treatment
of this problem.

It is important to note that even at very small values of $\gamef$
($\sim 0.01$), the fast drop of the pdf at low densities is still
observed, due to the factor $e^{(\gamef-1)s}$ in the exponential in
eq.\ \ref{eq:PDFgne1}, which in turn implies that there is always a
range of $s$-values in which the pressure is
not negligible in the hydrodynamic case, for any $\gamef$. This leads to
the
speculation that the pdf for Burgers flows, which are strictly
pressureless, should exhibit power laws at both large and small densities. 
This speculation is also verified numerically (\cite{PVS98}). Thus, the
Burgers case appears to be singular, not being the limit of
hydrodynamic flows as $\gamef \rightarrow 0$, at least as far as the
pdf is concerned.

\section{Conclusions}\label{conclusions}

In this Chapter we have discussed a scenario in which turbulence plays
a fundamental role in the production and determination of interstellar
cloud properties. Large-scale turbulent modes 
intervene in the former, while small-scale modes seem to  participate
in determining cloud scaling relations. ISM features that appear
naturally in our simulations are the phase-like appearance (a
consequence of turbulent density fluctuation production together with
an effective polytropic exponent $\gamef<1$), density and magnetic
field topologies and field strength ranges, and the velocity
dispersion-size relation and cloud mass spectrum. However, the
suggestions are made that the Larson (1981) density-size scaling
relation may be an artifact of surveys which do not integrate for long
enough times, and that the logatropic equation of state
(\cite{liz_shu89}) is not verified in highly dynamic situations.

Given the possible nature of clouds as turbulent density fluctuations,
their production in polytropic flows was also
discussed. The density jump across shocks and a criterion for the
collapse of these fluctuations were advanced. Finally, a model for the
probability density function of the fluctuations was described, which
satisfactorily explains the pdf shapes observed in isothermal, polytropic
and Burgers cases. Future work in this area will address the fully
thermodynamic and magnetic cases, aiming at explaining other forms of the
pdf, which have been observed numerically, but not produced by the model.

\begin{acknowledgments}
We gratefully acknowledge fruitful conversations with Annick Pouquet
and Susana Lizano. The
numerical simulations have been performed on the Cray Y-MP 4/64 of
DGSCA, UNAM, and the Cray C98 of IDRIS, France. This
work has received partial funding from grants UNAM/DGAPA IN105295 and
UNAM/CRAY SC-008397 to E.\ V.-S and from the PCMI National Program
of C.N.R.S. to T.P.
\end{acknowledgments}

\end{document}